# Dissipative-centrifugal instability of the Burgers vortex core and cyclone-anticyclone vortex asymmetry

S.G. Chefranov [1], A.G. Chefranov [2]

[1] A.M. Obukhov Institute of Atmospheric Physics RAS, Moscow, Russia
**schefranov@mail.ru**,

[2] Eastern-Mediterranean University, Famagusta, North Cyprus
Alexander.chefranov@emu.edu.tr

## Abstract

A new exact solution of the hydrodynamics equations is obtained in the form of the Burgers vortex generalization (BVG) accounting for the medium rotation as a whole and linear uniform friction. The dissipative-centrifugal nonlinear instability (DCI) condition, associated with some super-threshold rotation angular velocity and nonzero friction, for solid-body rotating core of BVG is obtained. The new effect of cyclone-anticyclone vortex asymmetry, arising during DCI of BVG vortex core, is shown to be similar to that observed for the atmospheres of high speed rotating planets and laboratory experiments.

## 1. Introduction

Modeling of the three-dimensional (3D) turbulent flows in laboratory and in computer simulations yield the vortex localization regions usually described by exact stationary solutions of the Burgers vortex (BV) type [1-4]. Stationary mode of such vortices (having a core with solid-body fluid rotation) is realized due to the balance between the 3D effect of the vortex stretching and volumetric viscosity forces hindering emergence of the enstrophy singularity in finite time in the unbounded space [1, 5, 6].

Up to now, BV is considered to be stable [2-4]. However, impact of the external friction as well as the rotation of the system as a whole on the BV stability has not been considered.

Here we obtain the BV generalization (BVG) which as BV is an exact solution of hydrodynamics equations but when the rotation of the system and uniform linear friction are taken into consideration. For the BVG solid-body core rotation instability conditions are also found herein. This instability, as we show, is related with the dissipative-centrifugal instability (DCI) [9, 10]. The DCI conditions for the BVG core velocity field are obtained that are realized only for the system with super threshold rotation angular velocity and effect of the linear uniform (external) friction. This result may be also interesting in relation with flow modes, emerging from coherent vortex structures modeling by a solid-body rotation, the frequency of which may be determined by uniform (external) friction [7,8].

Thus, DCI theory of the Lagrange fluid particles developed in [9, 10] is generalized here for the BVG core Euler's velocity field instability realization.

We also show that under BVG core DCI, chiral vortex symmetry is violated. The corresponding cyclone-anticyclone asymmetry emerges manifesting itself in the domination of the anticyclonic, or vice versa, cyclonic vortices depending on the system rotation frequency range (or respective Rossby numbers). The comparison of DCI theory conclusions with the data of laboratory and field experiments observing [11-17] cyclone-anticyclone vortex asymmetry is made.

The rest of the paper is organized as follows. In Section 2, we obtain a new solution of the Navie-Stokes equations in the form of stationary BVG and give the results which were obtained in [9] on the DCI condition to compare with the conclusions of our paper on the BVG Euler characteristics instabilty. In Section 3, the nonlinear (Sections 3.1-3.4) and linear (Section 3.5) instability of BVG are considered. In Section 4 the comparison of the theory and observation data is considered. Section 5 concludes the paper.

## 2. Burgers vortex generalization

The Navier-Stokes equations have the next form in a rotating reference frame that in the axial-symmetric case in coordinates $(r, z, \varphi)$ when the linear uniform (external or Eckman) friction also takes into consideration [10, 18]:

$$\frac{\partial V_r}{\partial t} + V_r \frac{\partial V_r}{\partial r} + V_z \frac{\partial V_r}{\partial z} - \frac{(V_\varphi + r\Omega)^2}{r} = -\frac{1}{\rho_0} \frac{\partial p}{\partial r} - 2\alpha V_r + \nu(\Delta V_r - \frac{V_r}{r^2}) \tag{1}$$

$$\frac{\partial V_\varphi}{\partial t} + V_r \frac{\partial V_\varphi}{\partial r} + V_z \frac{\partial V_\varphi}{\partial z} + \frac{V_r V_\varphi}{r} + 2V_r \Omega = -2\alpha V_\varphi + \nu(\Delta V_\varphi - \frac{V_\varphi}{r^2}) \tag{2}$$

$$\frac{\partial V_z}{\partial t} + V_r \frac{\partial V_z}{\partial r} + V_z \frac{\partial V_z}{\partial z} = -\frac{1}{\rho_0} \frac{\partial p}{\partial z} - g + \nu \Delta V_z; \Delta = \frac{1}{r} \frac{\partial}{\partial r} r \frac{\partial}{\partial r} + \frac{\partial^2}{\partial z^2} \tag{3}$$

$$\frac{\partial V_z}{\partial z} + \frac{1}{r} \frac{\partial (rV_r)}{\partial r} = 0 \tag{4}$$

Here $\nu$ is the kinematic viscous coefficient for a rotating with the frequency $\Omega$ incompressible fluid flat layer of the thickness $L$. We also use in (1), (2) the following definition of the uniform friction coefficient $\alpha = \frac{\sqrt{\Omega \nu}}{L}$ (when $\Omega = 0$, usually $\alpha = \frac{\nu}{L^2}$ [18]).

The system (1)–(4) in the semi-infinite region $z \geq 0$ has the following exact solution corresponding to the BVG:

$$V_r = \frac{dr}{dt} = -r\gamma; V_z = \frac{dz}{dt} = 2z\gamma , \tag{5}$$

$$V_\varphi = r\frac{d\varphi}{dt} = Br + \frac{\Gamma \exp(-2t\alpha)(1 - \exp(-r^2\gamma/2\nu))}{r} . \tag{6}$$

In (6) we have $B = \frac{\Omega \gamma}{\alpha - \gamma}$. When in (6) $\alpha = 0$, the solution (5), (6) exactly coincides with Burgers vortex (BV) when BV is considered in a rotating reference frame (with $B = -\Omega$ in (6)).

For the limit $t\alpha \to \infty$ in (6), the second term tends to zero, and, from (6), one gets
$$V_\varphi = Br . \tag{7}$$

Formulas (5), (7) also, as (5), (6), give an exact stationary solution of the hydrodynamics equations (1) - (4) [19].

For non-zero $\Omega$ and $\alpha$, the following relationship for the coefficient $\gamma$ in (5) holds $\gamma = \dfrac{\alpha B}{B+\Omega}$.

After dimensionless parameters introducing, it takes the form:

$$\gamma_{01} = \frac{\gamma}{\Omega} = \frac{\beta_0}{(1+\beta_0)\sqrt{\text{Re}}}, \beta_0 = \frac{B}{\Omega}, \text{Re} = \frac{\Omega L^2}{\nu} = \frac{1}{\alpha_1^2}, \alpha_1 = \frac{\alpha}{\Omega}.$$

When $\gamma < 0$ (for $\beta_0 < 0; |\beta_0| < 1$), in (5), radial motion of the Lagrange particles is directed from the reference frame origin.

For the BVG core corresponding to (5), (7), the pressure distribution is as follows [9]:

$$\frac{p_0}{\rho_0} = const - gz + \frac{\Omega^2}{2}\left[r^2\varepsilon_0 - 4z^2\gamma_{01}^2\right] |\varepsilon_0| = \frac{\Omega_0^2}{\Omega^2}$$

$$\varepsilon_0 = (\beta_0+1)^2 - \gamma_{01}^2(\beta_0) + \frac{2}{\sqrt{\text{Re}}}\gamma_{01}(\beta_0) \qquad (8)$$

In (8), the sign of $\varepsilon_0$ defines cyclonic (for $\varepsilon_0 > 0$), or anti-cyclonic (for $\varepsilon_0 < 0$) type of the radial pressure field distribution.

For a given value of $\varepsilon_0$ from (8), we get:

$$\beta_0 = -1 + \left[\frac{\varepsilon_0 \text{Re} - 1 + \sqrt{(\varepsilon_0 \text{Re} - 1)^2 + 4\text{Re}}}{2\text{Re}}\right]^{\frac{1}{2}}. \qquad (9).$$

Cyclonic type circulation orientation with $\beta_0 > 0$ in (9) takes place for $\varepsilon_0 > 1$. For the anticyclonic type circulation with $\beta_0 < 0$, from (9) we get the condition $\varepsilon_0 < 1$.

Note that according to [9] the DCI takes place when in (9) the following inequalities hold: $\beta_0 < 0; 1 + \beta_0 > 0; \Omega > |B|$.

The value of the right-hand side in (9) defines the Lagrange particle angular rotation velocity and the following relationship holds (from the determination $V_\varphi = r\dfrac{d\varphi}{dt}$):

$$\frac{d\varphi}{d\tau} = \beta_0; \tau = t\Omega, \qquad (10)$$

where the function $\beta_0(\text{Re}, \varepsilon_0)$ is represented in (9) for arbitrary positive parameters $\text{Re}$ and $\varepsilon_0$.

Under DCI condition $\beta_0 < 0$ the violation of the chiral vortex symmetry (see [9]) is exhibited in the realization only of the anti-cyclonic rotation of the Lagrange fluid particles (when $\dfrac{d\varphi}{d\tau} < 0$ condition holds in (9), (10)). This result may have connection to the observed cyclone-anticyclone vortex asymmetry in the systems with sufficiently fast rotating fluid and effects of the external uniform linear friction. When the DCI condition holds [9] for the Lagrange fluid particle radial coordinate, according to (5), we have the following exponential instability of the equilibrium state at the reference frame origin:

$$r(\tau) = r(0)\exp(\lambda_0 \tau);$$

$$\lambda_0 = -\frac{\beta_0}{(1+\beta_0)\sqrt{Re}} = \frac{1}{\sqrt{Re}}\left[\frac{1}{\sqrt{2}}(1-\varepsilon_0 Re + \sqrt{(\varepsilon_0 Re-1)^2 + 4Re})^{1/2} - 1\right] > 0 \qquad (11)$$

In (11), exponent $\lambda_0 = \lambda_0(Re, \varepsilon_0)$ tends to zero when Re>>1 and Re<<1 and has maximum for some finite Reynolds number, Re, that complies with the data of laboratory experiment [16] (see Fig. 1).

Vice versa, for the cyclonic rotation with $\beta_0 > 0$ (i.e. for $\varepsilon_0 > 1$ in (9)), in (10) according to (11), a particle equilibrium at the reference frame origin is exponentially unstable [9]. It is what constitutes the mechanism of violation of the cyclone-anticyclone vortex symmetry related to the realization of the Lagrange fluid particles DCI.

In the next section, we show that instability of the stationary mode (5), (7) has the same DCI mechanism providing now Euler velocity field (5), (7) instability yielding change of the dynamic mode.

## 3. Instability of the BVG core velocity field

### 3.1. The equations for disturbance of velocity and pressure fields
Let the non-stationary velocity and pressure disturbance fields are as follows:

$$V_{r1} = -\Omega\gamma_1(t)r; V_{z1} = 2\Omega\gamma_1(t)z; V_{\varphi 1} = \Omega\beta_1(t)r;$$

$$\frac{p_1}{\rho_0} = \frac{\Omega^2 \varepsilon_1(t)}{2}(r^2 + z^2\varepsilon) \qquad (12)$$

where three new unknown independent dimensionless functions $\gamma_1(t), \beta_1(t), \varepsilon_1(t)$ are introduced. In (12), for simplicity, is assumed that the dimensionless parameter, $\varepsilon$, is not depending on time and is arbitrary in sign and value. For $\varepsilon = 0$, in (12), we have uniform vertical pressure disturbance distribution corresponding to the quasi-two-dimensional solid-body rotation (see [7]).

Substituting the sum of the disturbance fields (12) (when (5), (7) are also taken in consideration) in the equation system (1)–(4), we get the following system of equations for three non-linear dimensionless functions, $\gamma_1(\tau), \beta_1(\tau), \varepsilon_1(\tau)$, of dimensionless time, $\tau = \Omega t$:

$$\varepsilon_1 = \frac{d\gamma_1}{d\tau} + \beta_1^2 - \gamma_1^2 + 2\beta_1(\beta_0 + 1) + \frac{2\gamma_1}{(1+\beta_0)\sqrt{Re}} \qquad (13)$$

$$\frac{d\beta_1}{d\tau} = 2\gamma_1\beta_1 + 2\gamma_1(\beta_0 + 1) - \frac{2\beta_1}{(1+\beta_0)\sqrt{Re}} \qquad (14)$$

$$(2+\varepsilon)\frac{d\gamma_1}{d\tau} = -4\gamma_1^2 + \varepsilon(\gamma_1^2 - \beta_1^2) - \frac{2\gamma_1(\varepsilon + 4\beta_0)}{(1+\beta_0)\sqrt{Re}} - 2\varepsilon\beta_1(\beta_0 + 1) \qquad (15)$$

Non-linear system (13)–(15) has a zero stationary solution corresponding to the undisturbed mode of the originally stationary BVG core (5), (7). System (13)–(15) except zero stationary solution has also other limit stationary modes tending to which already yields BVG core (5), (7) instability.

**3.2. Stationary and non-stationary solutions of (14), (15) at $\varepsilon = 0$**
Let us consider an example $\varepsilon = 0$ of an exact general stationary solution of the system (13)-(15), on the base of which we make conclusion on the conditions of the BVG core (5), (7) instability with respect to arbitrary by amplitude disturbance:

$$\beta_{1s} = -\frac{2\beta_0(1+\beta_0)}{1+2\beta_0}; \gamma_{1s} = -\frac{2\beta_0}{(1+\beta_0)\sqrt{\text{Re}}} = 2\lambda_0; \varepsilon_{1s} = -\frac{4\beta_0(1+(1+\beta_0)^2 \text{Re})}{(1+\beta_0)(1+2\beta_0)^2 \text{Re}}, \qquad (16)$$

where the functions $\beta_0$ and $\lambda_0$ are from (9) and (11) respectively.

Also for this example $\varepsilon = 0$, the general non-stationary solution of the system (14), (15) for any initial data is as follows:

$$\gamma_1(\tau) = \frac{2\lambda_0 \gamma_1(0) e^{4\lambda_0 \tau}}{\gamma_1(0)(e^{4\lambda_0 \tau} - 1) + 2\lambda_0};$$

$$\beta_1(\tau) = f(\tau)\left[-\frac{\beta_1(0)}{2\lambda_0} + \int_0^\tau d\tau_1 \frac{f_1}{f}\right];$$

$$f = -[2\lambda_0 + \gamma_1(0)(e^{4\lambda_0 \tau} - 1)]\exp(\frac{2\lambda_0 \tau}{\beta_0});$$

$$f_1 = \frac{4\lambda_0(\beta_0 + 1)\gamma_1(0)e^{4\lambda_0 \tau}}{2\lambda_0 + \gamma_1(0)(e^{4\lambda_0 \tau} - 1)}; \int_0^\tau d\tau_1 \frac{f_1}{f} = -4\lambda_0 \gamma_1(0)(1+\beta_0)I(\tau), \qquad (17)$$

$$I(\tau) = \frac{1}{4\lambda_0}\int_1^{\delta(\tau)} du \frac{u^{-\frac{1}{2\beta_0}}}{(2\lambda_0 + \gamma_1(0)(u-1))^2}, \delta(\tau) = \exp(4\lambda_0 \tau)$$

where $\beta_0$ is given in (9), and $\lambda_0$ is from (11).

### 3.3 The solution (17) for the cyclone type initial condition with $\gamma_1(0) > 0$.

That corresponds to the initial disturbance with vertical and radial structure, which usually is typical for vortices with cyclonic circulation orientation. It however does not impose any constraints on the selection of this or that circulation orientation sign that allows studying cyclonic as well as anti-cyclonic circulation of the stationary mode on stability.

The case of the initial cyclonic circulation with $\beta_0 > 0$ according to (17) and (11) leads to the negative $\lambda_0 < 0$ that provides exponential damping of the disturbances in time. It complies with the known data on Burgers vortex stability [1-4].

That is why, let us consider further with the help of the solution (17) stability of the stationary solution (5), (7) (i.e. solid-body BVG core rotation) that corresponds to anti-cyclonic orientation when $\beta_0 = -|\beta_0| < 0$.

Let also in (17) inequality $\lambda_0 > 0$ holds (for $\varepsilon_0 < 1$ with $-1 < \beta_0 < 0$, see (9), (11) and $\beta_0 = -|\beta_0|$) when for the flow mode (5), (7) DCI conditions (see [9]) are also take place. Then non-stationary solution (17) describes transition from (5), (7) to a new stationary mode (16) in the limit $\tau \gg 1$, thus yielding instability of the stationary mode (5), (7) of velocity field.

Actually in the limit $\lambda_0 \tau \gg 1$ according to (17) instead of (5), (7), a new stationary mode is realized, $\beta_s = \frac{|\beta_0|}{1-2|\beta_0|} > |\beta_0|$, corresponding also to the solid-body rotation, but with new frequency that is larger that initial frequency in (7). That rotation has cyclonic orientation only under the following condition:

$$|\beta_0| < \frac{1}{2} \text{ ( or in the dimensional form: } |B| < \frac{\Omega}{2}). \qquad (18).$$

Such a situation when from the originally anti-cyclonic vortex emerges more intensive cyclonic vortex corresponds to the noted in a number of works cyclonic-anti-cyclonic vortex asymmetry with cyclonic vortices prevailing [13]. It corresponds also to the observed in nature and laboratory experiment [17] when cyclonic vortex generation is following after appearing of an initial anti-cyclonic vortex mode.

Note that for $|\beta_0| = \frac{1}{2}$, reaching of the limit stationary mode in (17) for $\gamma_1(\tau) \to \gamma_{1s}$ is combined with linear in time increase of the tangential disturbance when $\beta_1(\tau) \to \frac{4|\beta_0|\tau}{\sqrt{Re}}$ and also in the result exactly cyclonic vortex is formed.

For $1 > |\beta_0| > \frac{1}{2}$, from (17), we get now exponentially growing solution for the velocity field tangential component disturbance that in the limit $\lambda_0 \tau \gg 1$ is as follows:

$$\beta_1(\tau) \cong C_0 \exp(2\lambda_0 \tau (2 - \frac{1}{|\beta_0|}));$$
$$C_0 = \gamma_1(0)\beta_1(0) + (1 - |\beta_0|)C_1, \tag{19}$$
$$C_1 = \int_1^\infty du \frac{u^{\frac{1}{2|\beta_0|}}}{(u - 1 + \frac{2\lambda_0}{\gamma_1(0)})^2}; \gamma_1(0) > 0$$

In (19) $\lambda_0 > 0$, where $\lambda_0$ depends on Re and $\varepsilon_0$ in the form (11) under DCI condition (see Fig. 1). Condition $\frac{1}{2} < |\beta_0| < 1$ holding corresponds to bounding not only from below (as required for DCI realization under conditions $B = -|B| < 0$, $\Omega > |B|$, or $|\beta_0| < 1$), but also from above of the system rotation frequency. It leads to the generalization of the DCI condition of [9] as follows:

$$|B| < \Omega < 2|B|, \tag{20}$$

According to (19), $\beta_1 \cong C_0 \exp(\lambda_1 \tau)$, where rate $\lambda_1$ of exponential growth in time of the Euler velocity field disturbance already does not coincide with the rate of exponential instability of Lagrange particles (11) though the former is defined via the latter as follows:

$$\lambda_1 = 2\lambda_0 (2 - \frac{1}{|\beta_0|}) = \frac{2(2|\beta_0| - 1)}{(1 - |\beta_0|)\sqrt{Re}} = \frac{2}{\sqrt{Re}} \left[ \frac{(1 - \varepsilon_0 Re + \sqrt{(\varepsilon_0 Re - 1)^2 + 4Re})^{\frac{1}{2}}}{\sqrt{2}} - 2 \right] \tag{21}$$

In the limit $\varepsilon_0 Re \gg 1$, value $\lambda_1 \cong \frac{2\sqrt{2}}{\sqrt{\varepsilon_0 Re}} \to 0$ in (21). Moreover, according to (21), instability takes place only for positive $\lambda_1 > 0$, when the Reynolds number exceeds the threshold value for sufficiently small parameter $|\varepsilon_0| = \frac{\Omega_0^2}{\Omega^2}$:

$$Re > Re_{th} = \frac{12}{1 - 4\varepsilon_0} > 0; \varepsilon_0 < \frac{1}{4} \tag{22}$$

Condition (22) is obtained from the inequality $|\beta_0| > \frac{1}{2}$, and for $\lambda_0 > 0$ holding it is required that $\beta_0 < 0$, which holds for $\varepsilon_0 < 1$.

Under condition (20) (see also (22) which is necessary for (20) realizing) from the initial anti-cyclonic stationary mode after its instability an intensive cyclonic vortex is developed for arbitrary small initial disturbances with the cyclonic circulation orientation (when $\beta_1(0) > 0$ and in (19), the constant $C_0 > 0$).

Vice versa, for realization of the resulting intensive anti-cyclonic vortex due to the instability under condition (20) (with $C_0 < 0$ in (19)) already it is required not simply any by amplitude initial anti-cyclonic disturbance but only sufficiently large by amplitude above-threshold disturbances meeting the following condition

$$|\beta_1(0)| > |\beta_{1th}| = \frac{(1-|\beta_0|)C_1}{\gamma_1(0)} \tag{23}$$

where $|\beta_0|$ and $C_1$ are functions of the Reynolds number $\mathrm{Re} = \frac{\Omega L^2}{\nu} = \frac{\Omega^2}{\alpha^2}$ and the parameter $\varepsilon_0 = \frac{\Omega_0^2}{\Omega^2}$ defined in (9) and (19). For example, in the limit $\mathrm{Re} \gg 1$, from (23) we get the following condition of developing of the anti-cyclonic vortex under instability condition (22):

$$|\beta_1(0)| > |\beta_{1th}| \approx \frac{\varepsilon_0 \sqrt{\mathrm{Re}}}{2(1-\sqrt{\varepsilon_0})} + O(\ln \mathrm{Re}) \tag{24}$$

Thus, we have got conditions of realization of cyclonic-anti-cyclonic vortex asymmetry that takes place under instability of the stationary vortex mode (5), (7), corresponding to the solid-body rotation of the BVG core. We show (for $\varepsilon = 0$ in (12) and (15)), that realization of instability of the BVG core under its initial stationary anticyclonic rotation with $-1 < \beta_0 < 0$ (i.e. under DCI condition) shall take place always under condition (22) and such instability leads to the development of exactly cyclonic vortex for arbitrary small disturbances with cyclonic circulation orientation.

Vice versa, only for the initially anticyclonic disturbances having above-threshold amplitude according to (23), (24) under condition (22) (or (20)) in the result of BV core instability, an intensive vorticity having anticyclonic circulation orientation is developed. These conclusions are related to the case when for the initial disturbance of radial and vertical velocity field components holds condition $\gamma_1(0) > 0$ (that corresponds to positivity of the vertical and negativity of the radial velocity field disturbance components that is typical videlicet for vortices with the cyclonic circulation orientation). Let us note that under condition $\gamma_1(0) > 0$ from (17), exponential instability follows of the mode (5), (7) of the BVG core solid-body rotation when it has cyclonic circulation orientation. It is also an example of the cyclonic-anti-cyclonic asymmetry since as it is shown above, under condition (22), the BVG core rotation with anti-cyclonic circulation orientation is unstable namely for $\gamma_1(0) > 0$.

**3.4 The solution (17) for the anticyclone type of initial condition with $\gamma_1(0) < 0$.**

From (17), also cyclone-anticyclone asymmetry follows that is exhibited in that the instability of the cyclonic (with $\beta_0 > 0$, when $\varepsilon_0 > 1$ and $\lambda_0 = -|\lambda_0| < 0$ according to (9) and (11)) BVG core (5), (7) is impossible for not too large values $\gamma_1(0) = -|\gamma_1(0)|$, satisfying the following condition $|\gamma_1(0)| < 2|\lambda_0|$.

Only in the opposite case with $|\gamma_1(0)| > |\gamma_{1th}| = 2|\lambda_0|$ the instability of the blowing type (when disturbances obtained infinite value during some finite time $t_0$ or dimensionless time $\tau_0 = t_0\Omega$) however, is possible. It's take place in finite time at the limit $\tau \to \tau_0^- = \frac{1}{4|\lambda_0|}\ln(\frac{1}{1-\frac{2|\lambda_0|}{|\gamma_1(0)|}})$, when from (17), we get $\gamma_1 \to -\infty$. Then for any initial disturbances $\beta_1(0)$ (independent from their sign and value), it tends exactly to the anti-cyclonic solid-body rotation when $\beta_1(\tau) \cong -(1+\beta_0)(1+O((\tau_0^- - \tau)\ln(\tau_0^- - \tau))) < 0$

On the other side, for the case of the anticyclone type rotation orientation (with $\beta_0 = -|\beta_0| < 0, \lambda_0 > 0$) of the BVG core in (5), (7) now for arbitrary by amplitude initial disturbances of $|\gamma_1(0)| > 0$ from (17) it follows an opportunity of the blowing instability. In this case for any initial disturbance $\beta_1(0)$ of the velocity field tangential component we get $\gamma_1(\tau) \to \infty$, and $\beta_1 \cong -(1-|\beta_0|)[1 + O((\tau_0^+ - \tau)\ln(\tau_0^+ - \tau))] < 0$ (under DCI condition $|\beta_0| < 1$) in the limit $\tau \to \tau_0^+ = \frac{1}{4\lambda_0}\ln(1 + \frac{2\lambda_0}{|\gamma_1(0)|})$.

Thus, also in the case under consideration of the negative initial values $\gamma_1(0) < 0$ cyclone-anticyclone asymmetry takes place. That asymmetry is exhibited in the absence of the threshold by amplitude of the disturbance amplitude $|\gamma_1(0)|$ only for the realization of instability for anticyclone type vortex mode in (5), (7). Besides that, for any initial disturbance $\beta_1(0)$ in the limit $\tau \to \tau_0^+$ (for $-1 < \beta_0 < 0$) as well as for $\tau \to \tau_0^-$ (when $\beta_0 > 0$) it takes place tendency only to anticyclone vortex mode with $\beta_1(\tau) < 0$ that makes anticyclone type vortex mode definitely distinguished with respect to a cyclonic one. Meanwhile, let us note that the DCI mechanism of the Lagrange particles defined in (11) not dependent explicitly on the Reynolds number (that anyway shall be finite) is found out to be different from the DCI mechanism of the Euler velocity field realized only for above-threshold Reynolds numbers according to the DCI condition (22).

**3.5 The linear instability theory for arbitrary value $\varepsilon$ in system (13)-(15).**

Let us consider also stability of the zero solution of the system (14), (15) in the linear approximation when in that system all non-linear terms can be neglected.

The rate of exponential evolution of small disturbances (having the form $\beta_1(t) = \beta_1(0)\exp(\lambda_2\tau), \gamma_1(\tau) = \gamma_1(0)\exp(\lambda_2\tau)$) is obtained from (14), (15) in the form (under condition $\varepsilon \neq -2$):

$$\lambda_2 = \frac{-2(\varepsilon + 2\beta_0 + 1) \pm 2\sqrt{(2\beta_0 - 1)^2 - \varepsilon(2+\varepsilon)(1+\beta_0)^4 \text{Re}}}{(2+\varepsilon)(1+\beta_0)\sqrt{\text{Re}}} \quad . \tag{25}$$

From (25), it follows that $\lambda_2 > 0$ and linear instability of zero solution of the system (14), (15) definitely takes place under condition $-2 < \varepsilon < 0$, when inequality $-1 < \beta_0 < 0$ holds, that corresponds to realization of DCI for $\varepsilon_0 < 1$ in (9). In that case, condition of instability of Lagrange particles when $\lambda_0 > 0$ in (11) is coincide with the condition of the Euler velocity field instability when $\lambda_2 > 0$ in (25).

For $\varepsilon > 0$, positivity of $\lambda_2$ in (25) also takes place not only under the same restriction on $\beta_0$ from (9) (i.e. for $-1 < \beta_0 < 0, \beta_0 = -|\beta_0|$), but if the following inequality holds

$$0 < \varepsilon < \frac{4|\beta_0|}{1 + (1-|\beta_0|)^4 \, \mathrm{Re}} \qquad (26)$$

In the case when in (25) $\beta_0 > 0, -2 < \varepsilon < 0; \varepsilon = -|\varepsilon|$, linear instability is possible under the following condition

$$2 > |\varepsilon| > \frac{4\beta_0}{1 + (1+\beta_0)^4 \, \mathrm{Re}} \,. \qquad (27)$$

For the special case $\varepsilon = 2$ instead of (25) we get

$$\lambda_2 = \frac{2\left[(1+\beta_0)^4 \, \mathrm{Re} - 2\beta_0 + 1\right]}{(2\beta_0 - 1)(1+\beta_0)\sqrt{\mathrm{Re}}} \,. \qquad (28)$$

According to (28), instability is absent when $\frac{1}{2} > \beta_0 > -1$, but holds when $\beta_0 > \frac{1}{2}$ and

$$\frac{(1+\beta_0)^4 \, \mathrm{Re}}{(2\beta_0 - 1)} > 1 \,. \qquad (29)$$

When considering inequality (29) it is necessary using (9) with parameter $\varepsilon_0$ such that

$$\mathrm{Re} > \mathrm{Re}_{th}(\varepsilon = -2) = \frac{5}{9(\varepsilon_0 - 9/4)} > 0 \qquad (30)$$

For the right inequality in (30) holding, it is necessary to have $\varepsilon_0 > \frac{9}{4}$.

It is important noting that, from (28), the linear instability also takes place for all Re when $\beta_0 < -1$ and DCI can't take place in (11) and we obtain a new type of instability for BVG.

When $\varepsilon = 0$ from (25) and (14) we have exact compliance with the instability conditions obtained on the base of the solution (17) of the nonlinear equations (14), (15).

## 4. Comparison with observation data

The condition (20) of realization of cyclone (or anticyclone – only under additional condition (23)) mode in the result of instability of the initial anticyclone type stationary mode gives the generalization to conclusions of the theory [9, 10] in the relation to the stability of the Euler flow characteristics corresponding to the BVG core.

Let us note that the condition (20) of nonlinear instability yields constraints on the rotation frequency of system both from below and above. And the ratio of the maximal and minimal allowed rotation frequencies is equal to two in condition (20). That relationship exactly coincides with an estimate obtained from the helicity balance equation in [20] and is close to the ratio of the threshold system rotation frequencies defining the mode boundaries of an anticyclone rotating jet (ARJ, for the rotation frequencies $0.055 \, rad/s < \Omega < 0.21 \, rad/s$) observed in the laboratory experiment [15]. In [15], in the experimental study of the air motion in the closed rotating vessel (heated from below), three different rotating frequency regions are distinguished: in addition to ARJ: the region of tornado-like vortices emerging (TLV, for the frequencies $\Omega > 0.21\,rad/s$), and turbulent rotating convection (TRC, for the frequencies $0.02\,rad/s < \Omega < 0.055\,rad/s$).

Fig. 1 shows the rate of exponential growth of the disturbances in dependence on the Reynolds number obtained on the base of (11), (21) and (25). Fig.1 also shows the experimental data [16] (for the dependence of the intensity of the vortex observed on the Reynolds number) obtained in the laboratory modeling of the tropic cyclone emergence. There is qualitative and quantitative agreement between the theoretical conclusions and experimental data [16] on the existence of the system rotation lower and upper threshold frequencies for realization of the vortex structure.

In [14], analysis of the cyclone-anticyclone observing frequency depending on the scale of the vortices is conducted. It was found out that only for the horizontal scales, $L > L_{th} = 800\,km$, vortices with anticyclone type circulation prevail, but for relatively small scales with L<750 km the prevailing vortex structures are with the cyclonic type circulation.

It complies with the noted here and in [9, 10] Lagrange particles DCI condition when according to (11), actually, takes place exponential instability of the zero state of equilibrium with anticyclone type rotation direction only for the frequencies of the disturbances less than the system rotation frequency, $B < \Omega$. For example, for the system rotation frequency corresponding to the daily period, T= 86400 sec, and for the character velocity scale, V=10 m/sec, we get condition, $L > L_{th} \approx TV = 864\,km$. This value nearly exactly complies with the threshold value $L_{th} = 800\,km$ obtained in [14] and provides satisfying of the DCI realization condition according to (11) and [9, 10]. Thus, the development of the very anticyclone type circulation with spontaneous violation of the chiral vortex symmetry stated here and in [9, 10] may give better understanding of the observed in [14] cyclone-anticyclone asymmetry formation mechanism.

## 5. Conclusion

Thus, we obtain the new conditions for BVG instability, which also give an opportunity to understand the cyclone-anticyclone asymmetry phenomena observed in nature and laboratory experiments.

Made herein consideration of the flow Euler characteristics instability generalizes conclusions [9, 10] in the form of the condition (20) under which an initial anticyclone type stationary vortex (the vortex core of BVG) becomes unstable for arbitrary small initial disturbances and in the result of this instability it turns to a vortex with exactly cyclonic circulation orientation.

In the observations [17], actually, on the background of the original anticyclone, usually a cyclonic vortex is formed rapidly. Noted in [13] domination of the cyclonic vortices also corresponds to the pointed mechanism of Euler characteristics DCI when the original anticyclone vortex is unstable with respect to the being formed on its basis cyclonic vortex. We note also that only for sufficiently large by amplitude disturbances satisfying condition (23) and under the Euler characteristics DCI condition (20), in the result of instability there can be formed an intensive vortex having already anticyclone circulation orientation. This is a possible reason also for the observing the large anticyclone vortexes not so frequently as small vortexes of cyclone type as it is obtained in analysis represented in [14].

It is interesting to obtain similar (as in this paper) stability conditions for the dissipative Sullivan type vortex generalization, the new solution for which is presented in [21].

The work was supported by the Russian Science Fundation № 14- 17- 00806P

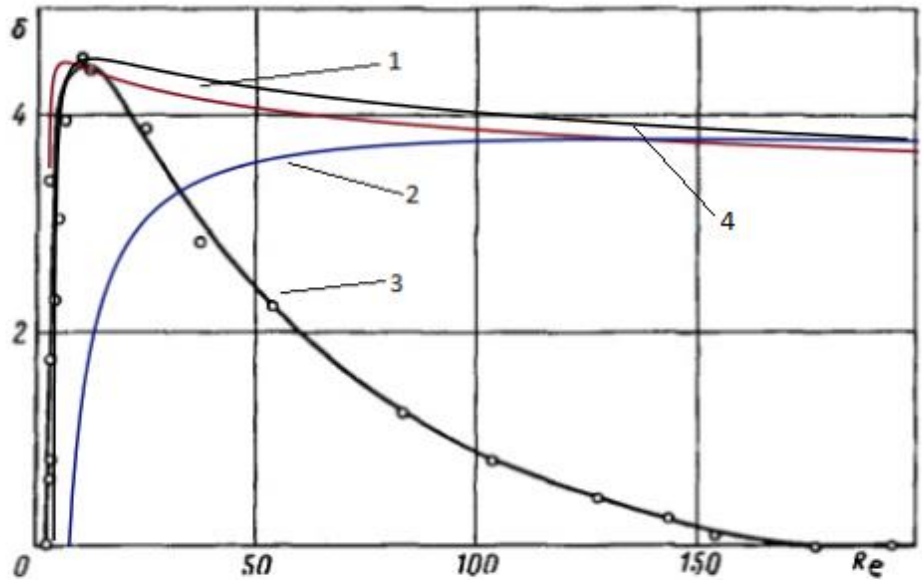

Fig.1. Comparison of the experimental data with the theoretic results in the case of $\varepsilon_0 = 0.05$. Curves 1, 2, and 4 are obtained using (11), (21) and (25), respectively (with introducing in the right-hand sides of the formulas of the numeric factor, 18). The curve 3 and respective points in its proximity correspond to the experimental data [16], where Re- the Reynolds number and $\delta$ - the parameter of vortex intensity.

## References

1. P. G. Saffman, Vortex Dynamics, Cambridge University Press, Cambridge, England, 1992